\documentclass[aps,preprint, prl, superscriptaddress, showpacs,nobibnotes]{revtex4-1}
\usepackage{graphicx}
\usepackage{bm,amsmath}
\usepackage{dcolumn}
\usepackage{natbib}

\begin{document}
\title{Transport Signatures of Fermi Surface Topology Change in BiTeI}

\author{Linda Ye}
\altaffiliation[Present Address: ]{Department of Physics, Massachusetts Institute of Technology, Cambridge, MA 02139, USA}
\affiliation{Department of Applied Physics, University of Tokyo, Tokyo 113-8656, Japan}
\author{Joseph G. Checkelsky}
\affiliation{Department of Physics, Massachusetts Institute of Technology, Cambridge, Massachusetts 02139, USA}
\author{Fumitaka Kagawa}
\affiliation{RIKEN Center for Emergent Matter Science (CEMS), Wako, Saitama 351-0198, Japan}
\author{Yoshinori Tokura}
\affiliation{Department of Applied Physics, University of Tokyo, Tokyo 113-8656, Japan}
\affiliation{RIKEN Center for Emergent Matter Science (CEMS), Wako, Saitama 351-0198, Japan}
\date{\today}

\begin{abstract}
We report a quantum magnetotransport signature of a change in Fermi surface topology in the Rashba semiconductor BiTeI with systematic tuning of the Fermi level $E_F$. Beyond the quantum limit, we observe a marked increase/decrease in electrical resistivity when $E_F$ is above/below the Dirac node that we show originates from the Fermi surface topology. This effect represents a measurement of the electron distribution on the low-index ($n=0,-1$) Landau levels and is uniquely enabled by the finite \emph{bulk} $k_z$ dispersion along the $c$-axis and strong Rashba spin-orbit coupling strength of the system. The Dirac node is independently identified by Shubnikov-de Haas oscillations as a vanishing Fermi surface cross section at $k_z=0$. Additionally we find that the violation of Kohler's rule allows a distinct insight into the temperature evolution of the observed quantum magnetoresistance effects.
\end{abstract}

%PACS, the Physics and Astronomy
%71.18.+y	Fermi surface: calculations and measurements; effective mass, g factor
%72.20.My	Galvanomagnetic and other magnetotransport effects
%71.70.Di	Landau levels
%71.70.Ej	Spin-orbit coupling, Zeeman and Stark splitting, Jahn-Teller effect

\maketitle

Dirac's matrix equation of relativistic electrons \cite{Dirac} has in recent years been found to be profoundly linked to the dynamics of electrons in solids. Application of this formalism has proven key in describing the intertwined (pseudo)spin degrees of freedom \cite{Geim-05,Fu-07,Bismuth}. One striking experimental implication of the linear dispersion is that in laboratory magnetic fields, it significantly enlarges low index Landau level energy separations compared to parabolic bands generated by a comparable tight-binding transfer integral, making emergent Dirac Fermion systems an ideal platform to study the quantum Landau level effects. Observations of this range from the discovery of Shubnikov-de Haas (SdH) oscillations in elemental bismuth \cite{SdH} to the first report of the quantum Hall effect at room temperature in graphene \cite{RTQH}. In this Letter we extend the quantum transport study of such Dirac structures to the critical point of a change in the \emph{bulk} Fermi surface topology. From a semiclassical point of view, it is understood that magnetoresistance (MR) is sensitive to the topology of given Fermi surfaces \cite{Pippard}: the magnetic field $B$ drives the electrons in orbits around the Fermi surface (FS), sensing its geometry and topology. We here describe the effect of the Dirac structure in this context of Fermi surface topology, which we observe as a magnetoresistance effect of pure quantum origin across the bulk Dirac node.

\begin{figure}
\includegraphics[width = \columnwidth]{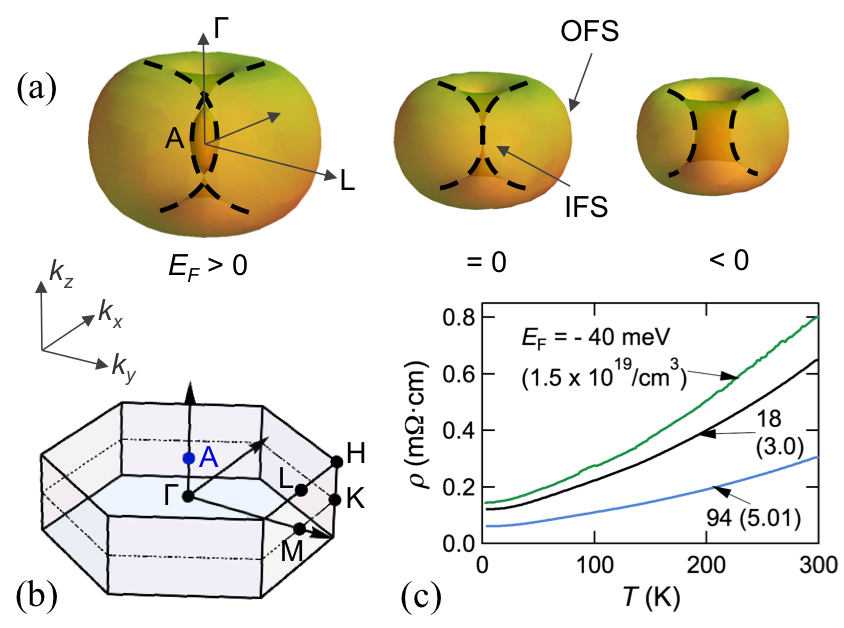}
\caption{\label{fig-1}(Color Online) (a) Schematic evolution of BiTeI Fermi surface (FS) from a spindle-torus ($E_F>0$) through horn-torus ($E_F=0$) to a ring-torus ($E_F<0$). Dashed lines define the IFS. (b) Location of FS in the Brillouin zone. (c) Representative $\rho(T)$ of samples with $E_F$ above (green), near (black) and below (blue) the Dirac point with Hall carrier density denoted in parenthesis. }
\end{figure} 

We study the system BiTeI that possesses a Dirac node generated by Rashba-type spin-orbit coupling $\bm{\alpha}\cdot(\bm{\sigma}\times\bm{k})$ \cite{Rashba,Rashba2}, where $\bm{\alpha}$ is the structure-specific Rashba parameter, $\bm{\sigma}$ and $\bm{k}$ are the spin and momentum operator, respectively. This layered semiconductor breaks inversion symmetry by crystallizing in the polar space group $P3m1$ so that the spin-orbit interaction takes the form of a Rashba term in the bulk 3D band structure. This has been confirmed by angle-resolved photoemission spectroscopy \cite{Ishizaka} and relativistic \emph{ab inito} calculations \cite{Bahramy-11}. In this system, both the conduction band minimum and the valence band maximum are located near the A point in the hexagonal prism-shaped Brillioun zone (Fig. 1(b)); taking A as the origin, the conduction electrons near the band edge can be described by the Hamiltonian
\begin{equation}\label{hamiltonian}
\mathcal{H}=\dfrac{\hbar^2k_z^2}{2m_z}+\dfrac{\hbar^2\bm{k}_{||}^2}{2m_0}+\bm{\alpha}\cdot(\bm{\sigma}\times\bm{k}_{||})
\end{equation}
where the momentum $\bm{k}$ is decomposed into $\bm{k}_{||}$ in the A-L-H plane and $k_z$ parallel to $\text{A}-\Gamma$ as well as $\bm{\alpha}$. The quasi-two-dimensionality of the crystal structure leaves $\mathcal{H}$ dominated by the in-plane components. Typical band parameters for BiTeI are $m_0=0.09m_e,|\bm{\alpha}|=3.85 \text{eV}\cdot\text{\AA}$ and $m_z/m_0\sim5$, as reported in photoemission \cite{Ishizaka}, optical \cite{Bordac-13} and transport \cite{Murakawa} studies. To emphasize the competition between the spin-splitting $\bm{\alpha}\cdot(\bm{\sigma}\times\bm{k}_{||})$ and kinetic energy $\hbar^2\bm{k}_{||}^2/2m_0$, $E_F$ is zeroed at the band crossing point at A, which is also the neutrality point of an effective Dirac Fermion with the Fermi velocity $v_F\approx 5.35\times10^5$ m/s \cite{Bordac-13}. Here the Dirac point defines a change of Fermi surface topology (Fig. 1(a)) from a spindle-torus ($E_F>0$) through a horn-torus ($E_F=0$)  to a ring-torus ($E_F<0$). The inner Fermi surface (IFS) and outer Fermi surface (OFS) are defined at each $k_z$ slice as shown in Fig. 1(a); the OFS is always electron-like while the IFS is bipolar depending on $E_F$ and $k_z$.

The simple band structure described by Eq. (\ref{hamiltonian}) and degenerate semiconductor nature (usually n-type)  of BiTeI allows a systematic exploration of this system upon doping, which readily transforms the FS progressively as depicted in Fig. 1(a). Single crystals of BiTeI were grown in a vertical Bridgman furnace, intentionally doped with Cu to improve electron mobility $\mu_e$ and vary the carrier density $n_e$  \cite{Taiwan-Cu}. MR and Hall effects are measured in four-probe configurations in a Quantum Design Physical Property Measurement System with $B$ applied along the polar $c$-axis. Compared with preceding studies \cite{SdH-1,Taiwan-Cu,Murakawa} our BiTeI samples exhibit relatively large $\rho^{\text{300K}}/\rho^{\text{2K}}=4\sim5.8$ (Fig. 1(c)) and enhanced low temperature $\mu_e$ typically between $800\sim3000$ cm$^2$/V$\cdot$s. $n_e$ varies within the range of $1.2\sim6\times10^{19}$/cm$^3$ for samples taken from the same ampoule, implying a composition gradient intrinsic to Bridgman growth \cite{Growth}. The microscopic role of Cu in improving the electronic quality of crystals is still being investigated. At the lowest temperature, we observe clear SdH oscillations over an extended range of $E_F$ as shown for selected samples in Fig. 2(a).

\begin{figure}
\includegraphics[width = \columnwidth]{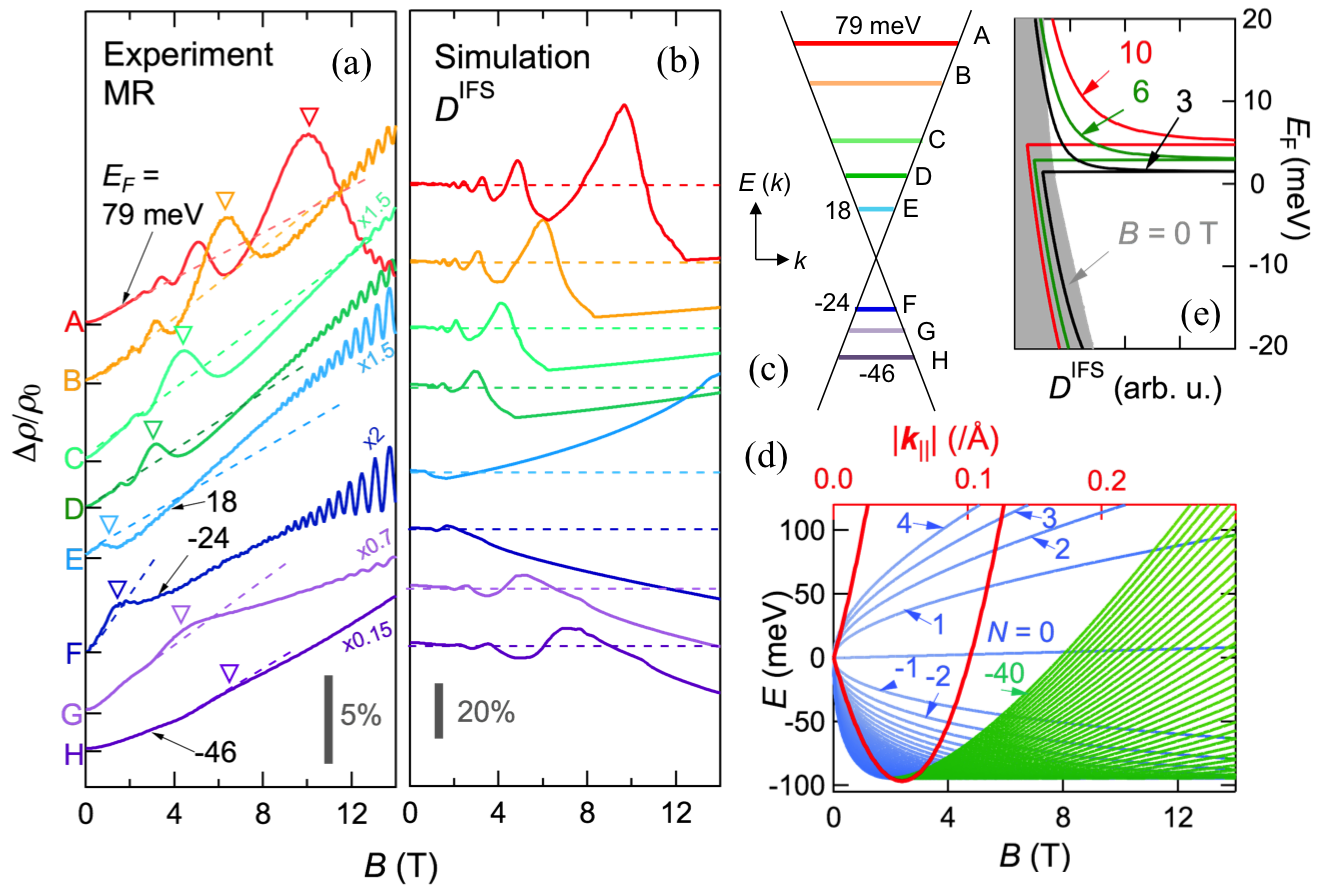}
\caption{\label{fig-2}(Color Online) (a) Normalized MR of samples A-H. Curves are scaled and offset to emphasize Landau level-related patterns. Triangles denote the IFS quantum limit and the dashed lines are extrapolation of the MR below the quantum limit. (b) Simulated IFS density of states $D^{\text{IFS}}$  for samples A-H. Curves are normalized and offset for clarity. (c) $E_F$ of samples A-H relative to Dirac point. (d) In-plane dispersion at $k_z=0$ when $B=0$ (upper axis) and selected Landau levels as a function of $B$ (lower axis). The colors blue and green represent respectively the IFS and OFS Landau levels. (e) Magnified $B$-evolution of $D^{\text{IFS}}$ near the Dirac point.}
\end{figure}

SdH oscillations is a common means to probe FS in metals \cite{Shoenberg}, whose frequency $f$ gives the extremal FS size $S_{\bm{k}}$ via $f=\hbar S_{\bm{k}}/2\pi e$. In Fig. 2(a) SdH of two different $f$ are traceable in all normalized MR curves ($\Delta\rho/\rho_0=(\rho(B)-\rho(0))/\rho(0)$): one consists of broad bumps starting from $B\sim$ 1 T, the other oscillates rapidly for $B\gtrsim10$ T. These represent two coexisting FS extremas of distinct size and are assigned to the IFS and OFS at $k_z=0$ in accordance with \cite{SdH-1,Taiwan-Cu,Murakawa}. $E_F$ is estimated \cite{e1} by comparing $f^{\text{OFS}}$ and $f^{\text{IFS}}$ to Eq. (\ref{hamiltonian}) with $m_0=0.095m_e,|\bm{\alpha}|=3.85$ eV$\cdot\text{\AA}$, which are determined below. Following the descending $f^{\text{OFS}}$ from 414 T (sample A) to 233 T (sample H), i.e., $E_F$ from 79 meV to -46 meV (see Fig. 2(c)), the field corresponding to the last IFS oscillation (defined as IFS quantum limit $B^{\text{QL}}$, denoted by triangles in Fig. 2(a)) approaches $B=0$ in samples A to E, then moves toward higher fields in samples F to H, consistent with the closing and opening of the inner FS pocket across $E_F=0$ in Fig. 1(a).

Beyond the IFS quantum limit, we observe distinctly different behavior for $E_F>0$ and $E_F<0$. It is most illuminating to contrast in Fig.2 (a) samples E ($18$ meV) and F ($-24$ meV) which bracket the Dirac point. In sample E, at $B>B^{\text{QL}}$ the MR grows rapidly above 2 T with a steeper slope than $B<B^{\text{QL}}$, while in sample F, the MR at $B\geq B^{\text{QL}}$ is strongly suppressed relative to $B<B^{\text{QL}}$. This distinction is systematically visible in samples B-D, G-H with higher $B^{\text{QL}}$. Apparently the MR beyond the QL provides a sensitive probe of the sign of $E_F$ and corresponding Fermi surface topology in BiTeI. This quantum transport behavior and its connection to FS topology are the main result of this Letter.

We suggest that the physical origin of this bifurcation is the quantum transport of the relevant Landau levels (LL) for the density of states (DOS) at $E_F$ \cite{Shoenberg}. Figure 2(d) shows the in-plane electronic states without (upper axis) and with (lower axis) $B$. The quantized LL energies are given by \cite{Rashba,Zeeman}
\begin{equation}\label{LL}
E_n=n\dfrac{\hbar eB}{m_0}\pm\sqrt{\dfrac{\hbar^2 e^2 B^2}{4m_0^2}+2n\dfrac{\alpha^2}{\hbar}eB}
\end{equation}
where $n$ is the LL index, including $n=0$ $(E_0=\hbar eB/2m_0)$. Selected LLs ($n=-40\sim4$) are shown in Fig. 2(d) and are divided into two groups colored blue and green, whose successive intersection with $E_F$ gives rise to the IFS and OFS oscillations, respectively. We note that because $E_n(B)$ for $n<0$ are non-monotonic, in increasing $B$ these LLs experience a transition from IFS-like to OFS-like. We illustrate this with the change in color of the negative LLs in Fig. 2(d). Blue (IFS) LLs resemble that of a Dirac Fermion with a slight $B$-dispersive $n=0$ LL due to the zero-point energy of the parabolic band with the effective mass $m_0$. Green (OFS) LLs can be seen to restrict the overall chemical potential within a weak variation. 

In contrast to this quantization in transverse $\bm{k}_{||}$, the longitudinal $k_z$ is unaffected, recalling a residual 1D contribution from every LL satisfying $E_n<E_F$ \cite{ReviewSdH}. As the OFS LLs are insensitive to the sign of $E_F$, we focus on the IFS to understand the observed MR behavior. The IFS density of states $D^{\text{IFS}}$ is given by: 
\begin{equation}\label{Sum}
D^{\text{IFS}}\bigg|_{E=E_F}=\sum\limits_{E_n<E_F}^{n\in \text{IFS}}\dfrac{\sqrt{2m_z}}{2\pi\hbar}\dfrac{1}{\sqrt{E_F-E_n}}\times\dfrac{eB}{2\pi\hbar}
\end{equation} 
with $E_n$ explicitly expressed in Eq. (\ref{LL}) and $E_F$ taken as $B$-independent for each sample. In Eq. (\ref{Sum}), $\sum\sqrt{2m_z}/2\pi\hbar\sqrt{E_F-E_n}$ counts the states at each cyclotron motion guiding center and $eB/2\pi\hbar$ counts the number of guiding centers. Note that we include the partial contribution of the lowest LL consistent with our definition of IFS as discussed above. The simulated $D^{\text{IFS}}$ for samples A-G from Eq. (\ref{Sum}) are displayed in Fig. 2(b) with each $E_n$ broadened with a Lorentzian 40 K wide.

Figure 2(b) reproduces the essential results in Fig. 2(a) including the bifurcation between samples with $E_F>0$ and $E_F<0$. This remarkable agreement suggests that the MR behavior reflects a quantum Landau level effect and is a measure of $D^{\text{IFS}}$. In Eq. (\ref{Sum}) the inverse square-root  $1/\sqrt{E_F-E_n}$ is most sensitive to the uppermost LL below $E_F$. Thus $D^{\text{IFS}}$ at $B>B^{\text{QL}}$ captures the difference between $E_0$ ($E_F>0$) and $E_{-1}$ ($E_F<0$) as a function of $B$. In the vicinity of the Dirac point, Fig. 2(e) shows a magnified view of $D^{\text{IFS}}$ with $B$ gradually turned on. At $E_F>0$, the contribution to $D^{\text{IFS}}$ from the weakly dispersive $n=0$ LL grows steadily, reflecting primarily the increasing capacity per LL proportional to $B$. At $E_F<0$, the suppression of $D^{\text{IFS}}$ is due predominantly to the rapid -$\sqrt{B}$ dispersion of $n=-1$ LL, which in the present $B$ range overwhelms the $B$-linear LL degeneracy term. 

The finite $m_z$ in the \emph{bulk} Rashba band structure is key to this effect by acting as a particle-hole symmetry (PHS) breaking term. In the context of band structure, in addition to the deformation of the Dirac node due to $m_0$, $m_z$ breaks PHS of the node by forcing both the in-plane electrons and holes to be electron-like in $k_z$. This allows qualitatively different MR properties of the Dirac node in BiTeI at $E_F>0$ and $<0$, in contrast to the nodes in graphene \cite{Geim-05} and Weyl semimetals \cite{Weyl}, where $E_F$ and $-E_F$ yields identical MR profile due to PHS. In the context of quantized Landau levels, $m_z$ breaks PHS when $E_F$ is exactly between two adjacent LLs of energies $E_a<E_b$: the inverse square root preferentially samples $E_a$ and thus enables measuring the dispersion of $E_a$ with $B$ in transport, as is the present case. This manner of PHS breaking is general for Landau quantization in 3D systems \cite{ReviewSdH} and has only recently been addressed in the context of the Nernst effect near the quantum limit in graphite \cite{Graphite}.

Beyond dimensionality and symmetry, the interconnection by the spin with a large OFS also distinguishes the transport of the Dirac node in BiTeI from that in graphene. Conventionally, it may be expected that the IFS and OFS would simply be additive in their contributions to the conductivity tensor ($\overset{\text{\tiny$\bm\leftrightarrow$}}{\sigma} = \overset{\text{\tiny$\bm\leftrightarrow$}}{\sigma}^{\text{OFS}} + \overset{\text{\tiny$\bm\leftrightarrow$}}{\sigma}^{\text{IFS}}$) so that $\Delta \sigma$ due to LL formation would follow $D^{\text{IFS}}$. The agreement of Fig. 2(a) and (b), however, shows empirically that $\Delta\rho\sim D^{\text{IFS}}$.  We hypothesize that this is caused by the large difference in size of the spin-polarized IFS and OFS, which becomes extreme near the Dirac point. Considering such a scattering phase space, for the IFS the inter IFS-OFS backscattering events dominate over the intraband backscattering. The major role of the IFS, among the total electrical current carried almost completely by the OFS, is then through $D^{\text{IFS}}$ modulations that affect the interband scattering rate $1/\tau_{\text{I-O}}$ (and thus $\rho$).

\begin{figure}
\includegraphics[width = \columnwidth]{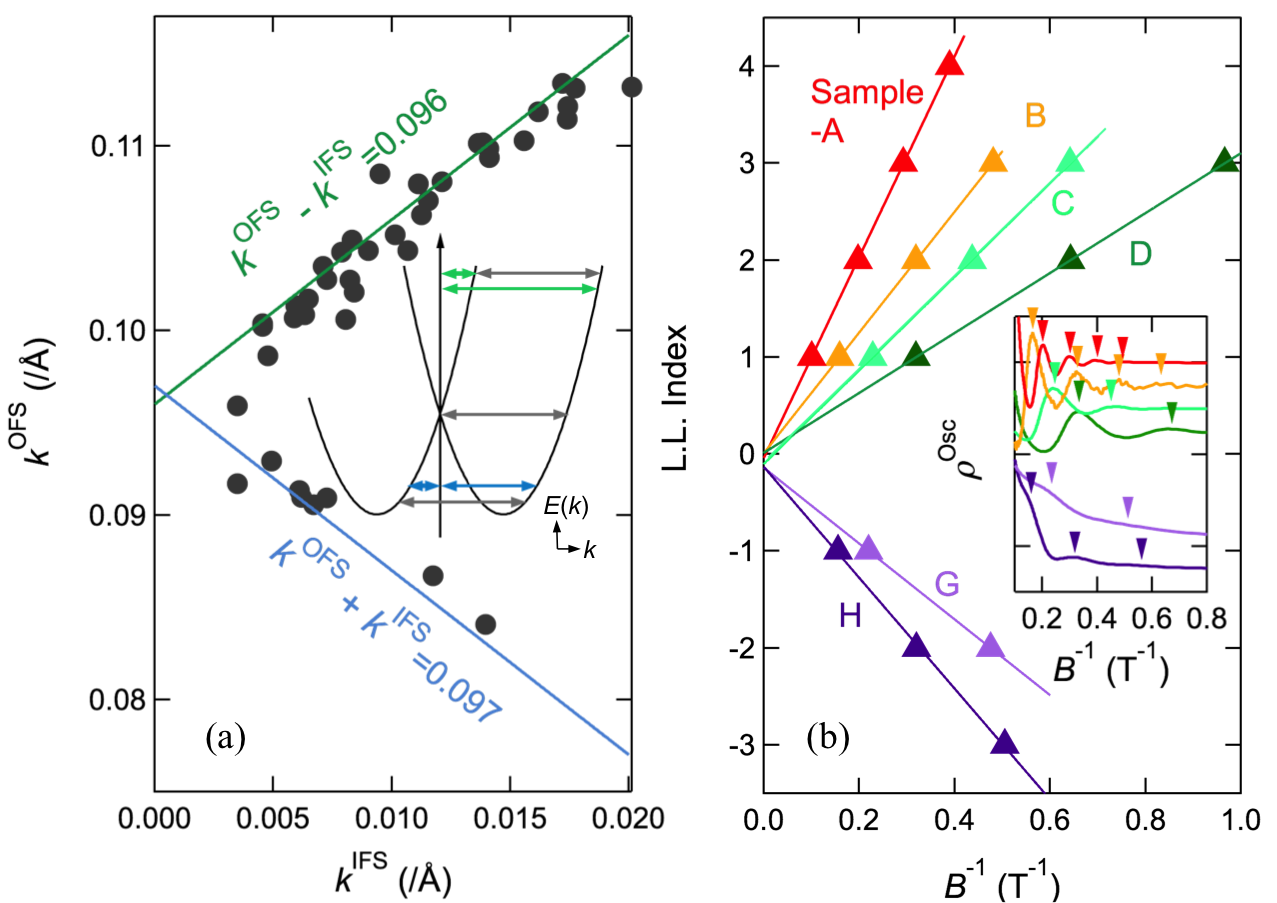}
\caption{\label{fig-3}(Color Online) (a) OFS and IFS radii at $k_z=0$ ($k_{\text{OFS}}$ and $k_{\text{IFS}}$) for all measured samples fitted with Eq. (\ref{V}). The Rashba bands are shown in the inset. (b) IFS index plot of samples A,B,C,D,G and H. The inset shows the oscillatory resistivity after background subtraction. }
\end{figure}

The above identification of FS topology is independently supported by the observed SdH oscillations, most unambiguously shown by comparing the FS radii $k^{\text{OFS}}$ and $k^{\text{IFS}}$ at $k_z=0$ as in Fig. 3(a). This captures the vanishing IFS at the Dirac point where $f^{\text{OFS}}\simeq$ 320 T, as expected from the relationship between the two coupled Fermi surfaces:
\begin{equation}\label{V}
\begin{split}
k^{\text{OFS}}-k^{\text{IFS}}&=2k_{\alpha}\quad \text{for}\quad E_F\geq0 \\
k^{\text{OFS}}+k^{\text{IFS}}&=2k_{\alpha}\quad \text{for}\quad E_F\leq0.
\end{split}
\end{equation}
The term $k_{\alpha}=m_0|\bm{\alpha}|/\hbar^2$ is the offset of band minimum generated by $\bm{\alpha}\cdot(\bm{\sigma}\times\bm{k}_{||})$, which performs a pure translation (gray arrow in Fig. 3(a) inset) between two spin branches on any $\bm{k}$-space cut through the $k_z$ axis. Eq. (\ref{V}) is a natural consequence of Eq. (\ref{hamiltonian}), confirming that this expression effectively describes the electrons near $E_F=0$. $k_{\alpha}$ is estimated to be $0.049\pm0.008$/$\text{\AA}$. This is consistent with $m_0=0.095m_e,|\bm{\alpha}|=3.85$ eV$\cdot\text{\AA}$ which we have used in the calculation above and is in reasonable agreement (within 6 \%) with previous reports \cite{Ishizaka,Bordac-13}, reassuring that the Cu-doping which improves the carrier mobility preserves the Rashba band structure. The IFS SdH senses both positive and negative low-index LLs across the Fermi surface topology change, as shown in the index plot Fig. 3(b).
  
\begin{figure}
\includegraphics[width = 0.9\columnwidth]{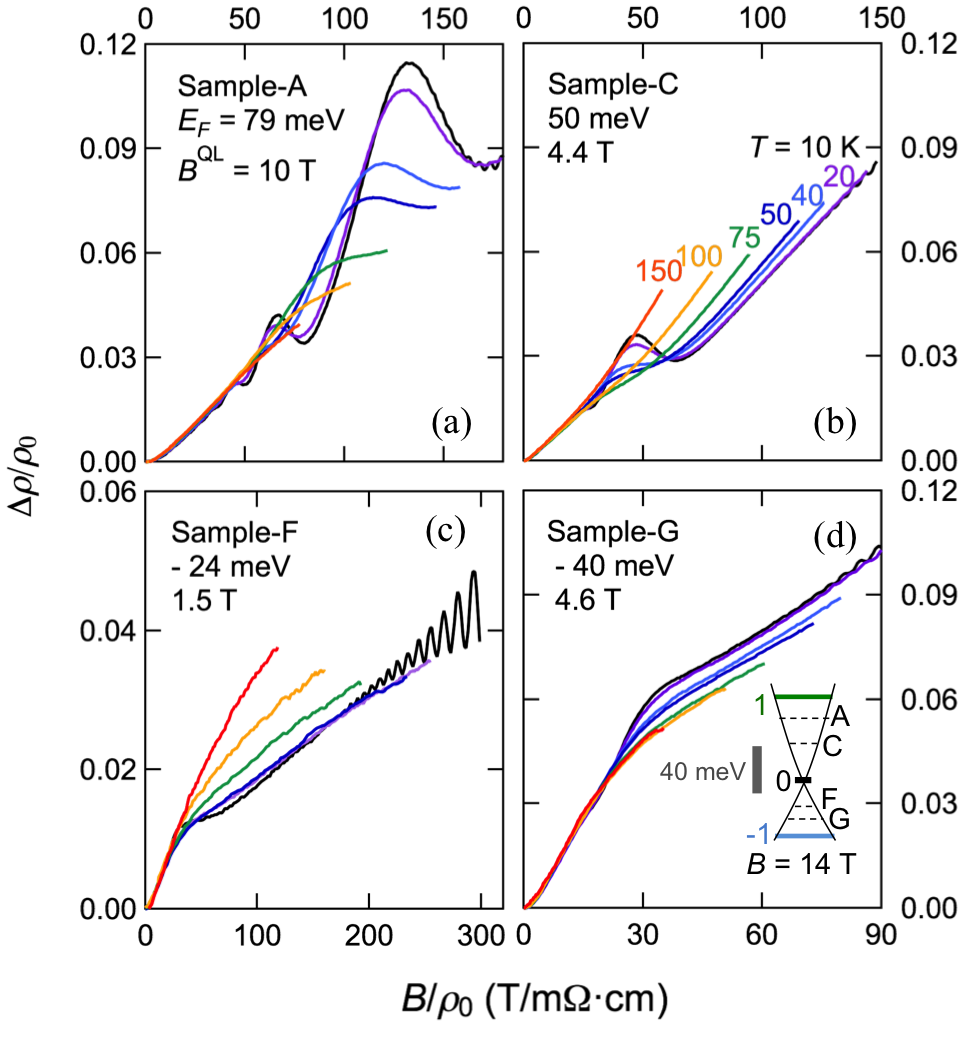}
\caption{\label{fig-4}(Color Online) (a)-(d) Temperature evolution of MR captured by the Kohler's plot of samples A,C,F,G. The inset of (d) shows the positions of $E_F$ at 14 T for each sample.}
\end{figure}

We next discuss the temperature ($T$) dependence of the observed quantum transport behaviors. The effect of $T$ on SdH oscillations is a useful probe of the Landau levels and associated carrier effective mass and scattering \cite{Shoenberg}. We have found that Kohler's rule \cite{Pippard}, which states that $\Delta\rho/\rho_0$ depends only on the Hall angle $\omega_c\tau$ and thus $B/\rho_0$, provides an incisive tool for isolating the LL contribution to transport across the QL. The Kohler's plot of MR curves taken at various $T$ of four different Fermi surfaces (samples A,C,F,G) are shown in Fig. 4. First, for $B<B^{\text{QL}}$, the MR curves of each FS fall onto a single trace despite oscillatory deviation, implying that the scattering is hardly affected by $B$ or $T$. Taking this collapsed curve as the background $\rho^{\text{Back}}$ onto which the SdH is superimposed assists the IFS index analysis in Fig. 3(b). We note that a previous work \cite{Taiwan-Cu} adopts a similar procedure in studying SdH in BiTeI. 

The validity of Kohler's rule within the QL restricts the most plausible cause of the violation (at high fields beyond the QL in Fig. 4(b)-(d)) to quantum effects. We suggest that these violations are consistent with the unconventional Dirac LLs, in particular the $n=0$ and $n=-1$.  MR of samples C and F are notably enhanced with elevated $T$, reflecting thermal excitation from the $n=0$ LL both upward and downward (see inset of Fig. 4(d)). In sample G, although the IFS SdH appears weaker compared to C or A, the suppressed Kohler's curve upon warming implies that the broad shoulder like feature in the MR is an intrinsic peak associated with the meeting of the coherent $n=-1$ LL with the $E_F$. Additionally, the contrary $T$-evolution of samples C and G despite their similar $B^{\text{QL}}$ demonstrates the asymmetry of quantum transport above and below the Dirac node. In this way it can be seen that Kohler's scaling brings out the LL effects beyond the QL which would otherwise be overlooked.

The uniqueness of BiTeI lies in the particularly large value of $\bm{\alpha}$ compared to other Rashba systems \cite{Ishizaka,Bahramy-11}. This strengthens the Dirac Fermion behaviors observed above and determines the large inter-LL scale: at 14 T $E_1-E_0\simeq 88$ meV while $E_0-E_{-1}\simeq71$ meV. Furthermore, the Dirac point is located $\sim$94 meV above the conduction band minimum, making $E_F < 0$ accessible at an appreciable doping level without entering the localization regime \cite{Mott}. We note that this $E_F<0$ regime in Rashba systems is of intense theoretical interest from the standpoint of enhanced superconducting instabilities \cite{SCinstability}, spin torque efficiency \cite{Torque} and superconductivity with peculiar symmetry \cite{SCSymmetry}. Our present transport identification of the $E_F<0$ FS topology thus paves the way for realizing the material host for these exciting proposals.

In conclusion, we have established the quantum magnetotransport properties of the Dirac node in BiTeI, which is markedly asymmetric for $E_F>0$ and $<0$. This effect is the result of a change in the \emph{bulk} FS topology. Kohler's scaling is shown to be an effective tool to analyze the unconventional low-index Landau levels. We predict the latter effect will be useful in the general study of band crossings in multi-band systems.

\begin{acknowledgements}
We thank S. Bord\'acs, B. J. Yang, V. Fatemi, J. D. Sanchez-Yamagishi and T. Ideue for fruitful discussions. This work was supported by JSPS Grant-in-Aid for Scientific Research(S) No. 24224009 and the Funding Program of World-Leading Innovative R\&D on Science and Technology (FIRST program) on ``Quantum Science on Strong Correlation" initiated by the Council for Science and Technology Policy, Japan.
\end{acknowledgements}

\pagebreak


\begin{thebibliography}{99}
\bibitem{Dirac} P. A. M. Dirac Proc. R. Soc. Lond. A \textbf{117}, 610 (1928)
\bibitem{Geim-05} K. S. Novoselov \emph{et al}., Nature \textbf{438}, 197 (2005)
\bibitem{Fu-07} L. Fu \emph{et al}.,  Phys. Rev. Lett. \textbf{98}, 106803 (2007); 
\bibitem{Bismuth} Z. Zhu \emph{et al}., Phys. Rev. B \textbf{84}, 115137 (2011)
\bibitem{SdH} L. Shubnikov and W. J. de Haas Leiden Comm. 207a, 207c, 207d, 210a (1930)
\bibitem{RTQH} K. S. Novoselov \emph{et al}., Science \textbf{315}, 1379 (2007)
\bibitem{Pippard} A. B. Pippard \emph{Magnetoresistance in Metals} (Cambridge University Press, Cambridge, England, 1989)
\bibitem{Rashba} E. I. Rashba, Sov. Phys. Solid State \textbf{2} 1109 (1960)
\bibitem{Rashba2} Y. A. Bychkov and E. I. Rashba J. Phys. C \textbf{17}, 6039 (1984)
\bibitem{Ishizaka} K. Ishizaka \emph{et al}., Nature Mater. \textbf{10}, 521 (2011)
\bibitem{Bahramy-11} M. S. Bahramy \emph{et al}., Phys. Rev. B. \textbf{84}, 041202(R) (2011)
\bibitem{Bordac-13} J. S. Lee \emph{et al}., Phys. Rev. Lett. \textbf{107}, 117401 (2011); S. Bord\'acs \emph{et al}., Phys. Rev. Lett. \textbf{111}, 166403 (2013)
\bibitem{Murakawa} H. Murakawa \emph{et al}., Science \textbf{342}, 1490 (2013)
\bibitem{Taiwan-Cu} C. R. Wang \emph{et al}.,  Phys. Rev. B \textbf{88}, 081104(R) (2013)
\bibitem{SdH-1} C. Martin \emph{et al}., Phys. Rev. B \textbf{87}, 041104(R); C. Bell \emph{et al}., Phys. Rev. B \textbf{87}, 081109(R) (2013); T. Ideue \emph{et al}., Phys. Rev. B \textbf{90}, 161107(R) (2014)
%\bibitem{Supp} Online Supplement Materials
\bibitem{Growth} D. T. J. Hurle \emph{Handbook of Crystal Growth 2} (North-Holland, Amsterdam, 1994)
\bibitem{Shoenberg} D.Shoenberg \emph{Magnetic Oscillations in Metals} (Cambridge University Press, Cambridge, England, 1984).
\bibitem{e1} We first determine the sign of $E_F$ by comparing $f^{\text{OFS}}$ to $m_0=0.095m_e,|\bm{\alpha}|=3.85 \text{eV}\cdot\text{\AA}$ with the critical point for FS topology $f^{\text{OFS}}_c = 314$ T. To better describe the IFS behavior that dominates the present $B$ range, the value of $E_F$ is adjusted to fit $f^{\text{IFS}}$ (IFS quantum limit) with the above parameter set.
\bibitem{Zeeman} The influence of the external Zeeman effect on the orbital motion is neglected because near the Dirac point the internal spin-orbit field $B_{\text{eff}}\simeq2m_0|\bm{\alpha}|/g\mu_B\hbar^2=3200$ T ($g=2$) is much larger than the applied $B$.
\bibitem{ReviewSdH} L. M. Roth and P. N. Argyres \emph{Semiconductors and Semimetals} \textbf{1} 159 (Academic Press, New York, 1966)
\bibitem{Weyl} H. B. Nielson and M. Ninomiya Phys. Lett. B \textbf{130} 389 (1983)
\bibitem{Graphite} Z. Zhu \emph{et al}., Nature Phys. \textbf{6} 26 (2010)
\bibitem{Mott} N. Mott \emph{Metal-Insulator Transitions (2nd edition)} (CRC Press, Cambridge, England, 1990).
\bibitem{SCinstability} E. Cappelluti \emph{et al}.,  Phys. Rev. Lett. \textbf{98}, 167002 (2007)
\bibitem{SCSymmetry} L. P. Gor'kov and E. I. Rashba Phys. Rev. Lett. \textbf{87}, 037004 (2001)
\bibitem{Torque} K. Tsutsui and S. Murakami Phys. Rev. B \textbf{86}, 115201 (2012)
\end{thebibliography}
\end{document}